\documentclass[runningheads]{llncs}
\usepackage[T1]{fontenc}
\usepackage{comment}
\usepackage{graphicx}
\usepackage{subcaption}
\usepackage{booktabs}
\usepackage{url}
\usepackage{hyperref}
\usepackage{color}

\usepackage{color}

\usepackage{enumitem}
\usepackage{multirow}
\usepackage[dvipsnames]{xcolor}
\definecolor{DarkGreen}{HTML}{006400} 
\usepackage{soul}

\begin{document}
\title{The Brazilian Vaccination Debate on YouTube: Topics, Perspectives, and Engagement Dynamics}

\titlerunning{The Brazilian Vaccination Debate on YouTube}

\author{
Matheus S. Azevedo\inst{1} \and
Geovana S. de Oliveira\inst{1} \and
Andrea Failla\inst{2,3} \and
Alexandre M. de Sousa\inst{1} \and
Fabricio Murai\inst{4} \and
Ana Paula C. da Silva\inst{5} \and
Carlos H. G. Ferreira\inst{6}
}
\authorrunning{M. S. Azevedo et al.}
%
\institute{
Universidade Federal de Ouro Preto, João Monlevade-MG, Brazil
\and
ISTI, National Research Council (CNR), Pisa-TOS, Italy
\and
University of Pisa, Pisa-TOS, Italy
\and
Worcester Polytechnic Institute, Worcester-MA, USA
\and
Universidade Federal de Minas Gerais, Belo Horizonte-MG, Brazil
\and
Universidade Federal de Ouro Preto, Ouro Preto-MG, Brazil
}
\maketitle

%
%
\begin{abstract}
Vaccination debates are central to online public health communication, as COVID-19 intensified disputes over scientific authority, institutional trust, and political identity. Yet studies often isolate semantic structure, stance, misinformation, and engagement, leaving their interplay over time poorly understood. We conduct a multilevel computational text analysis based on language models applied to 1.27 million Brazilian YouTube comments from 2018 to 2024, using what is, to our knowledge, the largest dataset of Brazilian vaccine discourse on the Web. We contrast producer framing in titles with audience discourse in comments, integrating Topic-derived themes with engagement metadata, conversational timing, stance-derived vaccine positions, and pre-pandemic, pandemic, and post-pandemic periods. Results show that COVID-19 dominates biomedical and informational themes in titles, whereas comments span personal health reports, vaccine effects, information credibility, conspiracy narratives, and political disputes. Health-related macro-topics dominate in scale and persistence, while conspiratorial and political themes are associated with faster interactions and a greater concentration of vaccine-opposing engagement. Post-pandemic activity remains centered on health experiences, vaccine effects, and information credibility, indicating no return to the pre-pandemic thematic configuration. By integrating semantic, interactional, stance, and temporal dimensions, this study shows how audiences reframe producer-framed health content and how vaccine controversies persist beyond the acute pandemic period.
\keywords{NLP \and Social Media \and Infodemic \and Vaccination \and YouTube}
\end{abstract}

%
%
\section{Introduction}
\label{sec:intro}

Vaccination is one of the most consequential public health interventions in modern history, contributing to the prevention and eradication of many infectious diseases worldwide~\cite{rodrigues2020impact}.
Despite its demonstrated efficacy, vaccination has also become a recurring object of public controversy~\cite{larson2016state,ebeling:2022,malagoli2021look}. The COVID-19 pandemic intensified this tension by placing vaccines at the center of public debate. In this context, vaccination debates moved beyond biomedical evidence, as vaccines became associated with disease prevention, institutional trust, political identity, and distrust toward scientific authority~\cite{henkel2023association,ebeling:2022,cinelli2020covid}. In COVID-19's aftermath, vaccination and viral diseases also became a \textit{total social fact} in the Maussian sense~\cite{mauss2014gift}, permeating public and private life and contributing to the redefinition of collective identities~\cite{labora2024social,henkel2023association,mesina2024whose}.

Brazil offers a relevant setting for studying these dynamics. The country has historically maintained one of the world's most comprehensive immunization programs~\cite{Silva:2020}, making vaccination a central component of collective health culture. At the same time, vaccine hesitancy has remained a persistent concern~\cite{Alsabban:2021,malagoli2021look,oliveira2026shapes}. Recent public disputes around health, science, and political authority further shaped how vaccination was discussed in Brazil, making the country an important case for analyzing vaccine-related discourse in a highly active digital media ecosystem.

Computational social scientists have increasingly used social media data to study how public controversies unfold online~\cite{cinelli2020covid,yin:2022,venancio2024evidencias}. However, existing studies often examine stance, misinformation, engagement, and polarization separately from the semantic organization of vaccine-related themes~\cite{ebeling:2022,malagoli2021look,de2022understanding,malagoli2021caracterizaccaob,melton:2022}. Much of this literature also focuses on English-speaking contexts, short observation windows, or COVID-19-specific debates~\cite{hwang:2022,son2025agenda,lindelof:2023}. As a result, we still know little about how thematic structures evolve across multiple immunization contexts and how they relate to engagement and vaccine-supporting or vaccine-opposing positions before, during, and after the COVID-19 pandemic.

A recent study by~\cite{oliveira2026shapes} addresses part of this gap by introducing a large-scale longitudinal dataset of vaccine-related YouTube comments in Brazil from 2018 to 2024 and by analyzing stance dynamics over time.
That study shows how vaccine-supporting and vaccine-opposing stances evolve across the Brazilian YouTube media ecosystem. However, it does not systematically examine the semantic and interactional contexts in which these stances are expressed. In particular, it remains unclear whether vaccine-supporting and vaccine-opposing positions are associated with specific themes such as vaccine effects, information credibility, conspiracy narratives, or political disputes. It also remains unclear whether different thematic contexts differ in engagement volume, conversational density, and temporal persistence. Given YouTube's broad reach in Brazil and its role in information consumption, including discussions around health and public policy~\cite{newman2025digital}, understanding these dynamics is especially relevant.

Building on this prior work, we shift the analytical focus from stance classification itself to the thematic and interactional environments in which stance-derived vaccine positions appear. Throughout the paper, we use the term \textit{stance-derived vaccine positions} to refer to comment-level labels indicating whether a comment is Favorable, Against, or Inconclusive toward vaccination. We use the term \textit{vaccine-related perspectives} in a narrower analytical sense, referring to how these positions are interpreted within semantic macro-topics, engagement patterns, and pandemic periods. Thus, we do not claim to reconstruct users' full motivations, identities, or belief systems. Instead, stance labels are used as an analytical layer to examine how vaccine-related positions are expressed and engaged with across thematic contexts over time.

To guide the analysis, we address the following research questions:
\begin{description}[leftmargin=1.2cm,labelwidth=1cm,labelsep=0.2cm]
    \item[\textbf{RQ1:}] What semantic themes structure vaccine-related video content and audience discussions on Brazilian YouTube, and how do they differ in engagement volume, density, and conversational reactivity?
    \item[\textbf{RQ2:}] How are stance-derived vaccine-supporting and vaccine-opposing positions distributed across semantic macro-topics, engagement patterns, and pre-pandemic, pandemic, and post-pandemic periods?
\end{description}

Our findings show that vaccine-related discourse on YouTube extends beyond biomedical concerns, as audience discussions reframe health-oriented video content into experiential, informational, conspiratorial, and political themes. Health-related discussions dominate in scale, whereas controversial themes are associated with faster interaction and a higher concentration of vaccine-opposing engagement. The pandemic amplified all thematic areas, while post-pandemic activity remained concentrated around health reports, vaccine effects, and information credibility. These results show how semantic structures, engagement, stance-derived vaccine positions, and temporality are jointly associated with the organization of online vaccination debates.

\section{Related Work}\label{sec:related}

Research on vaccine-related social media discourse shows that online platforms shape public health communication, vaccine hesitancy, misinformation, and political contestation~\cite{yin:2022}.
Prior studies report that anti-vaccine content can gain high visibility and interaction on YouTube, where platform affordances may intensify contested engagement~\cite{lahouati:2020,song:2017,miyazaki:2024,melton:2022}.
In Brazil, vaccination debates reflect a historically strong immunization culture and recent disputes over public health, science, and political authority, making the country a relevant case for studying vaccine controversies in digital media~\cite{locatelli:2022,ebeling:2022}.
However, much of this work focuses on engagement, misinformation, or polarization patterns, leaving less explored the semantic narratives through which vaccine debates are organized.

Topic modeling is widely used to characterize online vaccine discourse, from LDA-based studies of safety concerns, side effects, and hesitancy~\cite{melton:2021} to BERTopic analyses of short, noisy social media content~\cite{grootendorst2022bertopic,son2025agenda,de2022understanding}.
These approaches support analyses of COVID-19 discourse, misinformation, public perception, and multilingual health debates~\cite{hristova2022media}.
Nevertheless, existing studies often examine specific vaccines, short observation windows, or isolated semantic structures, limiting our understanding of how vaccine-related narratives evolve across multiple immunization contexts and how they connect to engagement and stance-derived vaccine positions.

Social media data have also been used for public health surveillance, supporting the detection of vaccine hesitancy, misinformation, public trust dynamics, and emerging health concerns~\cite{hwang:2022,huang:2024,yin:2022}.
In the Brazilian context, studies of YouTube comments and WhatsApp groups show that large-scale user discussions can reveal public perceptions, health-related controversies, and misinformation narratives~\cite{locatelli:2022,dias2024analise}.
Yet, semantic structure, engagement dynamics, temporal change, and vaccine-related positions are commonly analyzed separately.
This separation makes it difficult to explain which narratives attract attention, which ones concentrate vaccine-supporting or vaccine-opposing engagement, and how these patterns change across pre-pandemic, pandemic, and post-pandemic periods.

\section{Methodology}
\label{sec:meth}

This section describes the data and analytical strategy used to answer our RQs.
\subsection{Data Source and Processing}
\label{subsec:data}

We analyze a longitudinal dataset of vaccine-related YouTube discussions in Brazil, introduced by~\cite{oliveira2026shapes}. The dataset was collected through the official YouTube Data API using vaccination-related keywords derived from the Brazilian National Immunization Program (PNI)\footnote{\url{https://www.gov.br/saude/pt-br/vacinacao/calendario}}. The keyword set covers vaccines from the national immunization schedule, including vaccine names, lexical variants, and expressions related to benefits, risks, and controversies surrounding vaccination.

The collection prioritized the Brazilian context by restricting searches to Portuguese-language content and Brazil as the target region. Automatic language detection was applied in the original dataset to retain Portuguese comments. The data span 2018 to 2024 and include comment text and metadata on videos, channels, users, timestamps, likes, replies, and comment-thread structure. The final corpus comprises 1,276,730 unique comments associated with 13,455 videos, 3,647 channels, and 549,500 distinct users.

\subsection{RQ1: Vaccine Themes and Engagement Dynamics}
\label{subsec:rq1_methods}

To identify thematic structures in video content and audience discussions, we use BERTopic~\cite{grootendorst2022bertopic}, which combines contextual embeddings, dimensionality reduction, density-based clustering, and class-based TF-IDF. We generate embeddings with BERTimbau Base~\cite{souza2020bertimbau}, a Transformer model trained for Brazilian Portuguese, reduce them with UMAP~\cite{mcinnes2018umap}, cluster them with HDBSCAN~\cite{mcinnes2017hdbscan}, and summarize topics using class-based TF-IDF keywords. We apply BERTopic at two levels. First, we model video titles to capture themes introduced by content producers. Second, we model individual comments to capture audience interpretation, disagreement, personal experiences, uncertainty, and reframing.

Different configurations are used for titles and comments to account for differences in corpus size, text length, and lexical diversity. Following BERTopic parameter-tuning guidelines\footnote{\url{https://maartengr.github.io/BERTopic/getting_started/parameter\%20tuning/parametertuning.html}}, parameters were adjusted to balance coherence, interpretability, and granularity. For video titles, UMAP uses 10 neighbors, 10 dimensions, minimum distance of 0.01, and cosine distance; HDBSCAN uses a minimum cluster size of 30 and 5 minimum samples. For comments, UMAP uses 15 neighbors and 10 dimensions with the same distance settings; HDBSCAN uses a minimum cluster size of 200 and 5 minimum samples to avoid excessive fragmentation. The vocabulary is limited to the 3,000 most frequent terms for titles and 15,000 for comments, excluding terms appearing in fewer than 1\% or more than 90\% of documents.

Because comment-level modeling produced fine-grained and partially overlapping topics, we manually aggregated semantically related comment topics into macro-topics. Two authors first inspected the 32 comment-level topics using top keywords, topic size, and representative comments, grouping them into macro-topics through inductive review. Based on this taxonomy, we created a codebook with definitions and inclusion/exclusion criteria. Two other authors then independently assigned each topic to one macro-topic, reaching Cohen's $\kappa = 0.78$. Disagreements were resolved through consensus after reviewing representative comments. As an additional sanity check, we sampled 30 comments from each macro-topic and verified their semantic consistency.

To examine engagement, we combine topic and macro-topic assignments with YouTube metadata. For each macro-topic, we compute the number of channels, videos, comments, distinct users, replies, and likes. We also measure inter-comment time and reply latency to distinguish themes that dominate in volume from those that generate faster or more reactive interactions.

\subsection{RQ2: Stance-Derived Vaccine Positions Across Pandemic Periods}
\label{subsec:rq2_methods}

For RQ2, we distinguish four analytical levels. \textit{Topics} are fine-grained BERTopic clusters extracted from video titles or comments. \textit{Macro-topics} are manually aggregated groups of semantically related comment topics. \textit{Stance labels} are comment-level classifications indicating whether a comment is Favorable, Against, or Inconclusive toward vaccination. \textit{Vaccine-related perspectives} refer to the interpretation of these stance labels within macro-topics, engagement patterns, and pandemic periods. We therefore use stance labels as operational indicators of expressed vaccine position, not as reconstructions of users’ broader beliefs, motivations, or identities.

We rely on the stance labels produced by the domain-specific model introduced by~\cite{oliveira2026shapes}. The model is based on Llama 3.1 8B with QLoRA fine-tuning and was adapted to Portuguese vaccine-related YouTube comments in Brazil. It classifies each comment into three mutually exclusive classes: \textit{Favorable}, when the comment supports vaccination or expresses trust in vaccines; \textit{Against}, when the comment rejects, questions, or delegitimizes vaccination; and \textit{Inconclusive}, when no explicit vaccine-related stance can be reliably inferred. In the original evaluation, the model achieved F1-scores of 0.88 for \textit{Against}, 0.91 for \textit{Favorable}, and 0.94 for \textit{Inconclusive}\footnote{\url{https://huggingface.co/gseovana/llama-vaccine-stance-ptbr-lora}}. We interpret these labels as comment-level signals of expressed position, rather than as stable user-level beliefs.

We divide the analysis into three periods following the World Health Organization timeline for COVID-19: pre-pandemic (Jan. 1, 2018–Mar. 10, 2020), pandemic (Mar. 11, 2020–May 4, 2023), and post-pandemic (May 5, 2023–July 1, 2024). For each macro-topic, stance-derived position, and pandemic period, we compute comments, replies, and likes. Because the \textit{Inconclusive} class does not express a clear vaccine-related position, comparisons of vaccine-supporting and vaccine-opposing engagement focus on the \textit{Favorable} and \textit{Against} classes. Percentages are calculated relative to the total \textit{Favorable} and \textit{Against} engagement within each macro-topic across all periods, allowing temporal comparison without larger macro-topics dominating the analysis.
\section{Results}
\label{sec:results}
This section presents the findings related to RQ1 and RQ2, as stated in Section~\ref{sec:intro}.

\subsection{RQ1: Vaccine themes and engagement dynamics}
\label{subsec:rq1}

Table~\ref{tab:video_topics} presents the 14 topics extracted from video titles. Together, these topics cover approximately 95\% of the videos in the corpus and provide a platform-level view of how content creators frame vaccine-related issues.

\begin{table*}[t]
\centering
\caption{Topics extracted from video titles.}
\label{tab:video_topics}
\renewcommand{\arraystretch}{-1.05}
\setlength{\tabcolsep}{4pt}
\resizebox{\textwidth}{!}{%
\begin{tabular}{crll}
\toprule
\textbf{ID} & \textbf{\#Videos (\%)} & \textbf{Name} & \textbf{Keywords} \\
\midrule
VT1  & 8,377 (62.26\%) & COVID-19: Vaccination Outcomes and Fatalities & dose, kids, years, deaths, take vaccine, coronavac \\
VT2  & 909 (6.76\%) & Infectious Diseases: Symptoms and Treatment & hepatitis, symptoms, h3n2, influenza, flu outbreak, treatment \\
VT3  & 597 (4.44\%) & Long COVID and Post-COVID Sequelae & sequelae, post covid, syndrome, smell, can cause, loss \\
VT4  & 421 (3.13\%) & COVID-19 New Variants: Truths and Myths & new variant, myths, truths, sars, cov, subvariant \\
VT5  & 386 (2.87\%) & HPV and Cervical Cancer: Vaccine and Treatment & hpv, cancer, cervical, vaccine, uterus, warts \\
VT6  & 349 (2.59\%) & Diseases, Epidemics and Vaccines & fever yellow, measles, flu spanish, tuberculosis, vaccine, mumps \\
VT7  & 290 (2.16\%) & Pulmonary Tuberculosis: Symptoms and Treatment & tuberculosis, treatment, pneumonia, cough, pain, lung \\
VT8  & 279 (2.07\%) & Bacterial and Viral Meningitis & meningitis, tetanus, bacterial, pneumonia, microbiology, meningococcal \\
VT9  & 238 (1.77\%) & Flu Outbreaks and Information & need know, everything, flu, cold, bird flu, outbreak \\
VT10 & 236 (1.75\%) & Poliomyelitis Resurgence Alert & poliomyelitis, risk return, paulo henrique, paralysis, vaccination, children \\
VT11 & 226 (1.68\%) & COVID-19: Causes of Death and Vaccination & now, people died, cause covid, pandemic, vaccinated, myocarditis \\
VT12 & 213 (1.58\%) & Varicella-Zoster Virus Diseases & herpes zoster, chickenpox, pain, varicella, shingles, dermatologist \\
VT13 & 147 (1.09\%) & Dengue and Endemic Diseases Surveillance & dengue, agent, epidemiological surveillance, endemic, zika, combat \\
VT14 & 141 (1.05\%) & Issuance of International Vaccination Certificate & certified international, vaccination covid, step, issue, travel, civp \\
\bottomrule
\end{tabular}%
}
\vspace{-0.2cm}
\end{table*}

The title-level analysis shows that vaccine-related content is centered on biomedical, informational, and disease-oriented themes rather than explicit ideological conflict. COVID-19 dominates this layer of the ecosystem: VT1, VT3, VT4, and VT11 together account for approximately 72\% of the videos assigned to title topics. Beyond COVID-19, the remaining topics address routine immunization and infectious diseases, including HPV, tuberculosis, meningitis, influenza, poliomyelitis, dengue, and zika. Administrative issues, such as international vaccination certificates, appear only marginally. Video title documents assigned to BERTopic's outlier cluster were not interpreted as substantive themes. Overall, this pattern suggests that video producers mainly introduce vaccination through public health and institutional frames.

User comments reveal a broader and more contested discursive environment. We identify 32 comment-level topics and manually group them into seven macro-topics (MTs), which account for approximately 74\% of all comments in the corpus (Table~\ref{tab:macro_topics}). 
These macro-topics show how audience discussions expand video-level biomedical framing into experiential, informational, conspiratorial, and political interpretations, reinforcing previous interpretations of vaccination as a \textit{total social fact} in the Maussian sense, where a single phenomenon permeates multiple spheres of social life.

\begin{table*}[t]
\centering
\caption{Macro-topics present in user comments.}
\label{tab:macro_topics}
\renewcommand{\arraystretch}{-1.05}
\setlength{\tabcolsep}{4pt}
\resizebox{\textwidth}{!}{%
\begin{tabular}{cllr}
\toprule
\textbf{MT} & \textbf{Name} & \textbf{Keywords} & \textbf{\#Comments (\%)} \\
\midrule
\multirow{2}{*}{MT1} &
  \multirow{2}{*}{Symptoms and Health Reports} &
  (01) never caught, pain, test, smell, taste, garlic &
  399,493 (31.29\%) \\
& & (07) herpes zoster, lots pain, already take, high pressure, lord jesus, felt bad & 34,563 (2.71\%) \\
\midrule
\multirow{6}{*}{MT2} &
  \multirow{6}{*}{Conspiracy Theories and Distrust} &
  (04) new order, end times, gonna die, god word, human race &
  37,514 (2.94\%) \\
& & (08) mark, beast, plague inc, die, hurting, scar & 28,685 (2.25\%) \\
& & (11) pharmaceutical industry, make money, new order, dose, hiv vaccine, to sell & 22,602 (1.77\%) \\
& & (12) guinea pigs, serving, experiments, russian roulette, want force, nobody obliged & 21,582 (1.69\%) \\
& & (19) created virus, laboratory, biological weapon, population reduction, bats, chinese & 14,441 (1.13\%) \\
& & (27) satan, world order, demons, virus, divine justice, want kill & 12,263 (0.96\%) \\
\midrule
\multirow{6}{*}{MT3} &
  \multirow{6}{*}{Vaccination: Aspects, Effects, and Questions} &
  (09) can give, years old, syrup, honey, age group, babies &
  28,317 (2.22\%) \\
& & (13) save lives, read comments, vaccines, idle chat, will take & 20,236 (1.58\%) \\
& & (15) side effect, no reaction, astrazeneca, chest pain, god comfort, took effect & 18,551 (1.45\%) \\
& & (16) post covid, sequelae, post vaccine, symptoms, experimental, can cause & 18,157 (1.42\%) \\
& & (20) rules, body, pain, mandatory take, right choice, want, force & 14,321 (1.12\%) \\
& & (22) live science, god science, live vaccine, save lives, natural selection, technology & 13,924 (1.09\%) \\
\midrule
\multirow{3}{*}{MT4} &
  \multirow{3}{*}{COVID-19: General Discussions} &
  (03) someone knows, died, strange, deputies, where vaccine, many days &
  47,689 (3.74\%) \\
& & (14) quarantine, enaldinho, leave home, stay home, world, days & 18,686 (1.46\%) \\
& & (18) bought, cost, pff2, gas, n95, free & 17,076 (1.34\%) \\
\midrule
\multirow{4}{*}{MT5} &
  \multirow{4}{*}{Communication and Sources of Information} &
  (05) dr j\'ulio, thanks dr, congratulations dr, explanations, clarifications, video &
  36,735 (2.88\%) \\
& & (10) congratulations, thanks doctor, castanhari, didactics, video, information & 23,633 (1.85\%) \\
& & (30) source, zap, conspiracy theories, telegram, source information, butantan & 8,839 (0.69\%) \\
& & (31) voices head, source voices, olavo carvalho, reliable sources, head pain, fever & 5,531 (0.43\%) \\
\midrule
\multirow{4}{*}{MT6} &
  \multirow{4}{*}{Political and Ideological Discussions} &
  (06) dictator, rotten media, terrorize, dumb people, trash globo, lives lost &
  35,673 (2.79\%) \\
& & (23) jair bolsonaro, government, guilt, president, re-elected, reason & 13,553 (1.06\%) \\
& & (24) little flu, athlete history, myth, common flu, small talk, little gun & 13,292 (1.04\%) \\
& & (28) very sure, agree, fully, bolsonaro president, lula president, honesty & 10,501 (0.82\%) \\
\midrule
\multirow{3}{*}{MT7} &
  \multirow{3}{*}{Miscellaneous} &
  (17) certainty, absolute, certainty vaccine, many doubts, shadow, works &
  17,204 (1.35\%) \\
& & (26) good news, finally, fake news, bad news, good vaccine, good people & 12,906 (1.01\%) \\
& & (29) marijuana, smoke, liquor, cigarette, smoker, kills covid, drink & 9,726 (0.76\%) \\
\bottomrule
\end{tabular}%
}
\vspace{-0.2cm}
\end{table*}

Tables~\ref{tab:macro_topics} and~\ref{tab:engag_metrics} show that comment macro-topics differ in semantic content and participation patterns. MT1 (Symptoms and Health Reports) is the dominant macro-topic, with 434,056 comments and the highest discussion density, 41.9 comments per video. This indicates that symptom reports, post-vaccination reactions, diagnoses, and personal illness experiences are frequent and concentrated within video comment sections. MT3 (Vaccination: Aspects, Effects, and Questions) adds a more vaccine-centered layer of experiential discussion, including efficacy, side effects, mandatory vaccination, and practical immunization questions. Although smaller than MT1, MT3 has the highest number of likes per comment (4.44), suggesting comparatively high engagement with vaccine-effect and practical-question comments.

The engagement profile changes for contested themes. MT2 (Conspiracy Theories and Distrust) is smaller than MT1 in absolute volume, but it still produces 21.4 comments per video, 0.43 replies per comment, and 4.25 likes per comment. This suggests that pharmaceutical distrust, experimentation narratives, and political or religious interpretations are associated with denser and more interactive discussions. MT6 (Political and Ideological Discussions) is also smaller in volume, but it is concentrated in fewer channels, spanning 4,882 videos across 873 channels. This corresponds to 5.6 videos per channel, the highest ratio among all macro-topics. Political interpretations of vaccination and COVID-19 therefore appear less broadly distributed across channels, but recurrent where they emerge. MT5 (Communication and Sources of Information) provides the clearest contrast. It appears across many channels and videos, reflecting the role of doctors, science communicators, explanatory videos, and source evaluation, but averages only 10.7 comments per video and has the lowest reply rate (0.27). Thus, informational content is widespread but less interactive, although these differences may partly reflect audience size and channel popularity rather than thematic effects.

\begin{table}[t]
\centering
\tiny
\caption{Engagement metrics for macro-topics. C/V = comments per video, C/U = comments per user, R/C = replies per comment, and L/C = likes per comment.}
\label{tab:engag_metrics}
\resizebox{0.8\textwidth}{!}{%
\begin{tabular}{crrrrrrrrrr}
\toprule
\textbf{MT} & \textbf{\#Ch.} & \textbf{\#Vid.} & \textbf{\#Com.} & \textbf{\#Users} & \textbf{\#Rep.} & \textbf{\#Likes} & \textbf{C/V} & \textbf{C/U} & \textbf{R/C} & \textbf{L/C} \\
\midrule
MT1 & 2,722 & 10,364 & 434,056 & 253,281 & 180,753 & 1,737,105 & 41.9 & 1.71 & 0.42 & 4.00 \\
MT2 & 1,292 & 6,398  & 137,087 & 95,008  & 59,203  & 583,106   & 21.4 & 1.44 & 0.43 & 4.25 \\
MT3 & 1,545 & 6,837  & 113,506 & 81,874  & 47,958  & 504,227   & 16.6 & 1.39 & 0.42 & 4.44 \\
MT4 & 1,112 & 5,590  & 83,451  & 61,445  & 37,962  & 338,838   & 14.9 & 1.36 & 0.45 & 4.06 \\
MT5 & 2,225 & 6,978  & 74,738  & 59,877  & 19,824  & 269,846   & 10.7 & 1.25 & 0.27 & 3.61 \\
MT6 & 873   & 4,882  & 73,019  & 52,864  & 29,200  & 285,838   & 15.0 & 1.38 & 0.40 & 3.91 \\
MT7 & 1,085 & 4,637  & 39,836  & 32,803  & 19,370  & 149,588   & 8.6  & 1.21 & 0.49 & 3.76 \\
\bottomrule
\end{tabular}%
}
\end{table}

\begin{figure*}[t]
    \vspace{-2mm}
    \centering
    \begin{subfigure}[t]{0.32\textwidth}
        \centering
        \includegraphics[width=\linewidth]{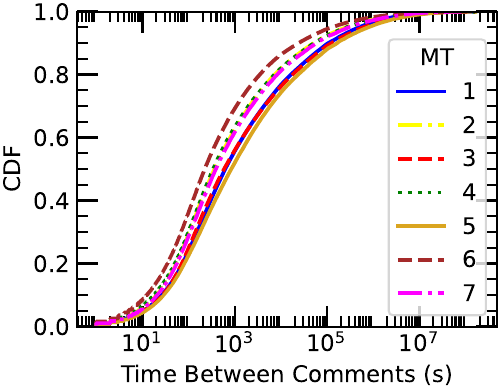}
        \caption{CDF of inter-comment time.}
        \label{subfig:cdf_inter_comment}
    \end{subfigure}
    \vspace{2cm}
    \begin{subfigure}[t]{0.32\textwidth}
        \centering
        \includegraphics[width=\linewidth]{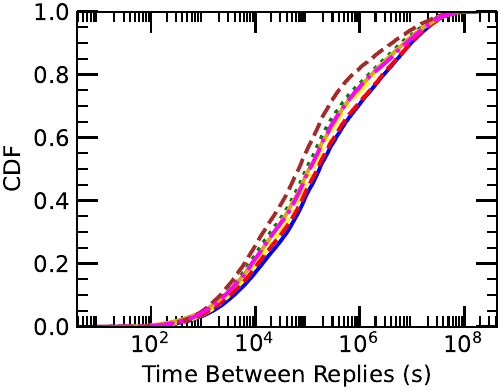}
        \caption{CDF of reply latency.}
        \label{subfig:cdf_reply_latency}
    \end{subfigure}
    \vspace{-2cm}
    \caption{Conversational dynamics across macro-topics. Time axes are in log scale.}
    \vspace{-0.2cm}
\label{fig:engagement_dynamics}
\end{figure*}

Figure~\ref{fig:engagement_dynamics} adds a temporal dimension to these engagement patterns by showing that thematic prevalence and conversational reactivity do not coincide. MT1 is the largest discussion space in volume and density, but it does not produce the fastest exchanges. Its median inter-comment time is 617 seconds and its median reply latency is 168,044 seconds ($\approx$47 hours), indicating that health-report discussions remain active over longer periods rather than unfolding as immediate exchanges. By contrast, contested macro-topics are associated with faster interaction despite their smaller scale. MT2 has a median inter-comment time of 360 seconds and a median reply latency of 125,660 seconds ($\approx$35 hours), while MT6 is the most reactive macro-topic, with the shortest median inter-comment time, 222 seconds, and a reply latency of 67,149 seconds ($\approx$19 hours). 

We formally assessed these differences using Kruskal--Wallis tests followed by Dunn post-hoc comparisons with Holm correction \cite{Hollander:2013}. Both temporal measures differed significantly across macro-topics: inter-comment time, $H(6)=11809.12$, $p<0.001$, and reply latency, $H(6)=285.13$, $p<0.001$. The key pairwise contrasts were consistent with the CDF patterns: MT2 and MT6 differed significantly from both MT1 and MT5 in inter-comment time and reply latency (all Holm-adjusted $p<0.001$). These patterns suggest that conspiratorial and political discussions are among the most conversationally intense parts of the debate, likely because they invite disagreement, contestation, and emotionally charged participation.

MT5 again contrasts with this pattern. Although communication and source-oriented discussions are widespread across channels and videos, they show slower interaction, including the highest median inter-comment time, 832 seconds. Together, Table~\ref{tab:engag_metrics} and Figure~\ref{fig:engagement_dynamics} show a clear divergence between scale, density, and reactivity: biomedical and experiential themes dominate participation volume, while conspiratorial and political themes are associated with faster conversational response.

\subsection{RQ2: Stance-derived vaccine positions across pandemic periods}
\label{subsec:rq2}

\begin{figure*}[t]
    \centering
    \begin{subfigure}[t]{0.45\textwidth}
        \centering
        \includegraphics[width=\linewidth]{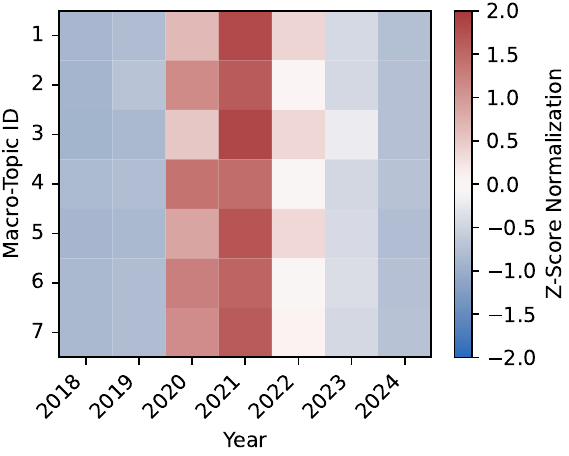}

        \caption{}        \label{subfig:temporal_evolution}
    \end{subfigure}
    \begin{subfigure}[t]{0.49\textwidth}
        \centering
        \includegraphics[width=\linewidth]{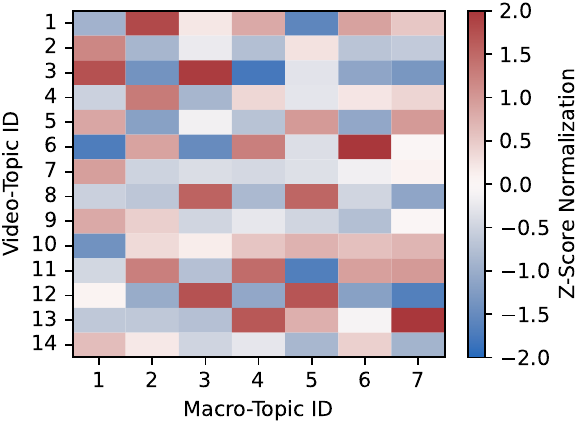}
        \caption{}
        \label{subfig:macro_x_video_topics}
    \end{subfigure}
    \begin{subfigure}[t]{0.70\textwidth}
        \centering
        \includegraphics[width=\linewidth]{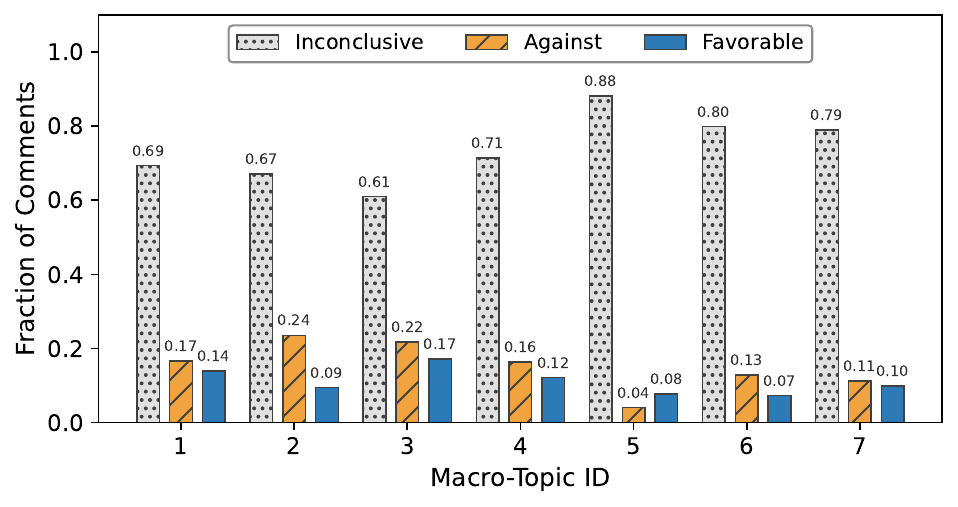}
        \caption{}
        \label{subfig:stance_by_topic}
    \end{subfigure}
 \caption{Macro-topic patterns: (a) temporal evolution, (b) video-topic association, and (c) distribution of vaccine-related positions.}
    \label{fig:temporal_stance}
    \vspace{-0.2cm}
\end{figure*}

We now examine how macro-topics evolve over time and how vaccine-supporting and vaccine-opposing engagement is distributed across pandemic periods. Following the operational definition in Section~\ref{subsec:rq2_methods}, stance labels are analyzed as comment-level indicators of expressed vaccine position.
Figure~\ref{subfig:temporal_evolution} presents the temporal evolution of macro-topic popularity using $z$-score normalization along the time axis, defined as $z=(x-\mu)/\sigma$, which allows comparison of relative topic intensity independently of comment scale. 

The temporal evolution reveals a marked increase across all macro-topics between 2020 and 2021. MT1, MT3, MT5, and MT7 peak in 2021, a period associated with the intensification of vaccination campaigns and the broader social experience of COVID-19 infections, sequelae, and vaccine reactions. This pattern is consistent with the content of these macro-topics: symptoms and health reports, vaccine effects and questions, communication and information sources, and heterogeneous discussions become more prominent when vaccination turns into a widespread everyday concern.

In contrast, MT2, MT4, and MT6 show high relative intensity already in 2020, indicating that conspiratorial, general COVID-19, and political discussions emerged early in the pandemic. This suggests that alternative explanations, institutional distrust, and political interpretations were part of the initial interpretive environment through which COVID-19 and vaccination were discussed on YouTube. In the post-pandemic period, MT1, MT3, and MT5 remain moderately active, suggesting that vaccine debates did not disappear after the acute pandemic phase, but shifted toward more persistent concerns about health experiences, vaccine effects, and reliable information.

Figure~\ref{subfig:macro_x_video_topics} shows that video-level themes and comment-level macro-topics do not map one-to-one. COVID-related video topics centered on deaths, vaccination outcomes, and causal attribution, especially VT1 (COVID-19: Vaccination Outcomes and Fatalities) and VT11 (COVID-19: Causes of Death and Vaccination), are overrepresented in MT2 (Conspiracy Theories and Distrust), MT4 (COVID-19: General Discussions), and MT6 (Political and Ideological Discussions). This suggests that videos framing COVID-19 through mortality, vaccine outcomes, and causal explanations often become entry points for distrust, broader pandemic disputes, and political contestation. By contrast, more clinical or disease-specific topics, such as VT3 (Long COVID and Post-COVID Sequelae), VT8 (Bacterial and Viral Meningitis), and VT12 (Varicella-Zoster Virus Diseases), are more associated with MT1 (Symptoms and Health Reports), MT3 (Vaccination: Aspects, Effects, and Questions), and MT5 (Communication and Sources of Information), suggesting that these videos more often sustain symptom reports, vaccine-related questions, and information-seeking. Thus, audience discussions reframe biomedical content differently depending on the video frame: COVID mortality and vaccine-risk topics are associated with macro-topics where vaccine-opposing comments are more common, while clinical disease topics are associated with experiential and informational engagement.

Finally, Figure~\ref{subfig:stance_by_topic} presents the distribution of stance-derived positions inferred from the comment-level classifier. 

With approximately 74\% of comments classified as \textit{Inconclusive}, this class predominates across all macro-topics. Consequently, comparisons between \textit{Favorable} and \textit{Against} should be interpreted as associations within the subset of comments expressing an identifiable stance, and are inherently conditioned on the predictions of the classifier. MT2 exhibits the strongest association with vaccine opposition: 24\% of its comments are classified as \textit{Against}, compared with 9\% as \textit{Favorable}. This difference is consistent with the semantic content of MT2, which includes themes of experimentation, pharmaceutical distrust, coercion, population control, and religious or apocalyptic explanations. A representative comment states: \textit{``We cannot be forced to take an emergency experimental vaccine. Our bodies should not serve as test subjects''}\footnote{For privacy reasons, all comments reported here were paraphrased and translated into English by the authors.}. This illustrates how vaccine opposition can be articulated through concerns about bodily autonomy, institutional distrust, and the framing of vaccination as experimentation.

However, vaccine-related position and semantic topic are not equivalent. Because BERTopic clusters comments by lexical and semantic proximity rather than by normative position, favorable comments can also appear within conspiracy-related discussions when they refute or contest misinformation. For instance, one comment states that \textit{``Only conspiracy theorists doubt the effectiveness of the vaccine and believe it contains chips, biological weapons, or is connected to 5G''}. This coexistence shows that comments with different vaccine-related positions may occupy the same semantic environment.

MT5 presents the opposite pattern. It is the only macro-topic in which favorable comments exceed opposing comments, with 8\% \textit{Favorable} and 4\% \textit{Against}. This is consistent with its emphasis on doctors, explanations, didactic videos, and source evaluation. MT1 and MT7 are more balanced, with MT1 containing 17\% \textit{Against} and 14\% \textit{Favorable} comments, and MT7 containing 11\% \textit{Against} and 10\% \textit{Favorable} comments. These patterns show that vaccine skepticism is not restricted to explicitly conspiratorial spaces. It also appears in personal health reports, heterogeneous discussions, and practical questions about vaccination.

\begin{table*}[t]
\centering
\caption{Vaccine-supporting and vaccine-opposing engagement across macro-topics and pandemic periods. Percentages represent the contribution of each position-period pair to the total vaccine-position engagement.}
\label{tab:stance_period_distribution}
\scriptsize
\setlength{\tabcolsep}{2.2pt}
\renewcommand{\arraystretch}{0.92}
\resizebox{\textwidth}{!}{%
\begin{tabular}{clrrr|rrr|rrr}
\toprule
\multirow{2}{*}{\textbf{MT}} & \multirow{2}{*}{\textbf{Position}} 
& \multicolumn{3}{c|}{\textbf{Pre-pandemic}} 
& \multicolumn{3}{c|}{\textbf{Pandemic}} 
& \multicolumn{3}{c}{\textbf{Post-pandemic}} \\
\cmidrule(lr){3-5} \cmidrule(lr){6-8} \cmidrule(lr){9-11}
& & \multicolumn{1}{c}{Comments} & \multicolumn{1}{c}{Replies} & \multicolumn{1}{c|}{Likes} 
  & \multicolumn{1}{c}{Comments} & \multicolumn{1}{c}{Replies} & \multicolumn{1}{c|}{Likes}  
  & \multicolumn{1}{c}{Comments} & \multicolumn{1}{c}{Replies} & \multicolumn{1}{c|}{Likes}  \\
\midrule
\multirow{2}{*}{1} 
& \textcolor{red}{\textit{Against}}   & 1,946 (1.47\%) & 566 (1.08\%) & 7,414 (1.38\%) & 63,336 (47.78\%) & 23,518 (44.73\%) & 265,227 (49.23\%) & 6,645 (5.01\%) & 1,861 (3.54\%) & 19,853 (3.69\%) \\
& \textcolor{DarkGreen}{\textit{Favorable}} & 1,507 (1.14\%) & 554 (1.05\%) & 7,553 (1.40\%) & 55,858 (42.14\%) & 24,907 (47.38\%) & 230,338 (42.75\%) & 3,277 (2.47\%) & 1,167 (2.22\%) & 8,359 (1.55\%) \\
\midrule
\multirow{2}{*}{2} 
& \textcolor{red}{\textit{Against}}   & 1,434 (3.17\%) & 265 (1.63\%) & 5,621 (2.97\%) & 27,830 (61.62\%) & 9,458 (58.27\%) & 122,705 (64.80\%) & 3,025 (6.70\%) & 781 (4.81\%) & 9,953 (5.26\%) \\
& \textcolor{DarkGreen}{\textit{Favorable}} & 798 (1.77\%) & 188 (1.16\%) & 3,887 (2.05\%) & 11,317 (25.06\%) & 5,251 (32.35\%) & 44,882 (23.70\%) & 763 (1.69\%) & 289 (1.78\%) & 2,300 (1.21\%) \\
\midrule
\multirow{2}{*}{3} 
& \textcolor{red}{\textit{Against}}   & 519 (1.17\%) & 143 (0.89\%) & 2,421 (1.29\%) & 20,618 (46.66\%) & 7,178 (44.60\%) & 98,210 (52.17\%) & 3,528 (7.98\%) & 882 (5.48\%) & 13,347 (7.09\%) \\
& \textcolor{DarkGreen}{\textit{Favorable}} & 558 (1.26\%) & 198 (1.23\%) & 1,189 (0.63\%) & 17,215 (38.96\%) & 7,078 (43.98\%) & 68,137 (36.19\%) & 1,754 (3.97\%) & 616 (3.83\%) & 4,946 (2.63\%) \\
\midrule
\multirow{2}{*}{4} 
& \textcolor{red}{\textit{Against}}   & 244 (1.02\%) & 71 (0.74\%) & 808 (0.92\%) & 12,373 (51.95\%) & 4,369 (45.65\%) & 49,019 (56.04\%) & 1,041 (4.37\%) & 281 (2.94\%) & 2,946 (3.37\%) \\
& \textcolor{DarkGreen}{\textit{Favorable}} & 199 (0.84\%) & 73 (0.76\%) & 516 (0.59\%) & 9,394 (39.44\%) & 4,576 (47.82\%) & 32,662 (37.34\%) & 568 (2.38\%) & 200 (2.09\%) & 1,513 (1.73\%) \\
\midrule
\multirow{2}{*}{5} 
& \textcolor{red}{\textit{Against}}   & 103 (1.17\%) & 12 (0.59\%) & 541 (1.60\%) & 2,611 (29.63\%) & 732 (36.09\%) & 9,738 (28.82\%) & 320 (3.63\%) & 77 (3.80\%) & 3,348 (9.91\%) \\
& \textcolor{DarkGreen}{\textit{Favorable}} & 192 (2.18\%) & 27 (1.33\%) & 441 (1.31\%) & 5,175 (58.72\%) & 1,092 (53.85\%) & 18,718 (55.40\%) & 412 (4.67\%) & 88 (4.34\%) & 998 (2.95\%) \\
\midrule
\multirow{2}{*}{6} 
& \textcolor{red}{\textit{Against}}   & 161 (1.10\%) & 46 (0.96\%) & 1,179 (2.29\%) & 8,428 (57.59\%) & 2,507 (52.47\%) & 30,352 (59.07\%) & 764 (5.22\%) & 161 (3.37\%) & 1,708 (3.32\%) \\
& \textcolor{DarkGreen}{\textit{Favorable}} & 74 (0.51\%) & 33 (0.69\%) & 326 (0.63\%) & 4,861 (33.21\%) & 1,930 (40.39\%) & 17,174 (33.43\%) & 347 (2.37\%) & 101 (2.11\%) & 640 (1.25\%) \\
\midrule
\multirow{2}{*}{7} 
& \textcolor{red}{\textit{Against}}   & 85 (1.01\%) & 26 (0.76\%) & 171 (0.49\%) & 3,961 (47.28\%) & 1,482 (43.17\%) & 15,965 (45.86\%) & 410 (4.89\%) & 130 (3.79\%) & 1,529 (4.39\%) \\
& \textcolor{DarkGreen}{\textit{Favorable}} & 92 (1.10\%) & 49 (1.43\%) & 125 (0.36\%) & 3,630 (43.33\%) & 1,657 (48.27\%) & 16,526 (47.47\%) & 199 (2.38\%) & 89 (2.59\%) & 499 (1.43\%) \\
\bottomrule
\end{tabular}%
}
\vspace{-0.2cm}
\end{table*}

Table~\ref{tab:stance_period_distribution} further connects stance-derived vaccine positions to engagement and time. For each macro-topic, the table reports the contribution of each stance-period pair to the total \textit{Favorable} and \textit{Against} engagement of that macro-topic across all periods. This normalization allows us to identify whether engagement is concentrated in favorable or opposing positions and whether this concentration changes across pandemic periods.

The results show that the pandemic concentrated most vaccine-position engagement across all macro-topics. However, the post-pandemic period suggests attenuation rather than disappearance. Although engagement declines after the 2020--2021 peak, it does not fully return to pre-pandemic levels. This is especially visible in MT1, MT3, and MT5. In MT1, post-pandemic \textit{Favorable} and \textit{Against} comments account for 7.48\% of the macro-topic total, compared with 2.61\% before the pandemic. In MT3, the difference is larger, increasing from 2.43\% before the pandemic to 11.95\% afterward. MT5 also remains comparatively active after the pandemic, particularly in likes, which rise from 2.91\% in the pre-pandemic period to 12.86\% in the post-pandemic period. These patterns suggest that, after the acute pandemic phase, vaccine debates persisted around personal health experiences, vaccine effects, and disputes over information credibility.

The distribution of stance-derived positions further shows that residual engagement differs across macro-topics. In MT2, opposing comments dominate pandemic engagement, accounting for 61.62\% of \textit{Favorable}/\textit{Against} comments, 58.27\% of replies, and 64.80\% of likes, compared with 25.06\%, 32.35\%, and 23.70\% for favorable comments. MT6 follows a similar pattern: opposing comments account for 57.59\% of comments, 52.47\% of replies, and 59.07\% of likes during the pandemic, while favorable comments account for 33.21\%, 40.39\%, and 33.43\%. Although these macro-topics decline after the pandemic peak, vaccine-opposing engagement remains proportionally stronger than favorable engagement. In MT2, for example, post-pandemic opposing comments and likes account for 6.70\% and 5.26\%, whereas favorable comments and likes account for 1.69\% and 1.21\%.

MT5 provides an important counterpoint. During the pandemic, favorable comments dominate this macro-topic, representing 58.72\% of \textit{Favorable}/\textit{Against} comments, 53.85\% of replies, and 55.40\% of likes, compared with 29.63\%, 36.09\%, and 28.82\% for opposing comments. This suggests that informational and source-oriented discussions were comparatively more aligned with vaccine-supporting engagement, likely because they include interactions with doctors, science communicators, explanatory videos, and evaluations of information credibility. However, the post-pandemic period shows a notable asymmetry in likes within MT5, as opposing comments account for 9.91\% of likes but only 3.63\% of comments. This suggests that, among comments that continued to receive engagement in MT5 after the pandemic peak, distrust-oriented contributions remained visible despite relatively low comment volume.

For MT1 and MT3, the position-period distribution shows that health-oriented discussions also contain vaccine-opposing comments and engagement, even though these macro-topics are not explicitly political or conspiratorial. During the pandemic, MT1 is relatively balanced in replies, with favorable comments accounting for 47.38\% and opposing comments for 44.73\%, while opposing comments account for a larger share of likes. After the pandemic, opposing engagement becomes more visible in both MT1 and MT3, especially in comments and likes. This suggests that personal health reports, side effects, and vaccine-related questions remained relevant entry points for skepticism after the pandemic peak.

Overall, RQ2 shows that engagement associated with vaccine-supporting and vaccine-opposing positions varies systematically across macro-topics and pandemic periods. The pandemic amplified all vaccine-related discussions, while the post-pandemic period reveals which forms of debate persisted. Conspiracy and political macro-topics declined after the pandemic peak but remained disproportionately associated with vaccine-opposing engagement. Communication and source-oriented discussions were more favorable during the pandemic, yet later showed signs of residual distrust. Health-report macro-topics remained mixed spaces where lived experience, uncertainty, and skepticism coexist.
\section{Conclusion and Future Work}
\label{sec:conclusion}

This paper investigated how vaccination debates evolved on Brazilian YouTube between 2018 and 2024 by integrating semantic analysis, engagement metadata, temporal segmentation, and stance-derived vaccine-position labels.
We used stance labels as an analytical layer to contextualize vaccine-supporting and vaccine-opposing comments within broader thematic and engagement patterns.

Our findings show that vaccine-supporting and vaccine-opposing engagement is unevenly distributed across thematic domains.
Health-related discussions dominate in volume and sustain long-term engagement, whereas political and conspiratorial narratives represent a smaller share of the debate but are associated with faster and more reactive interactions, while concentrating vaccine-opposing engagement.
The longitudinal analysis further suggests that the COVID-19 period was associated with a substantial reconfiguration of vaccine-related discussion on YouTube. Although activity peaked during the pandemic, post-pandemic discussions remained focused on vaccine effects, health experiences, and information credibility, suggesting that vaccination debates did not return to their pre-pandemic thematic configuration.

This study has limitations.
It focuses exclusively on YouTube and does not capture interactions across other platforms or offline contexts.
Topic modeling and manual macro-topic aggregation simplify complex discursive structures and remain partially dependent on qualitative interpretation.
Furthermore, automated vaccine-position classification may introduce residual uncertainty.
In addition, our use of the term vaccine-related perspectives is operationally bounded by stance labels.
The \textit{Favorable}, \textit{Against}, and \textit{Inconclusive} categories capture explicit comment-level positioning toward vaccination, but they do not fully represent users' broader motivations, identities, epistemic beliefs, or political orientations.
Therefore, our analysis should be interpreted as descriptive associations rather than evidence of causal relationships.
Future work could extend this analysis through cross-platform studies, multimodal approaches incorporating video content, network-based investigations of information diffusion, stability analyses across multiple BERTopic configurations, and predictive or causal modeling to better explain the mechanisms underlying the observed thematic and engagement patterns.

%
%
\bibliographystyle{splncs04} 
\bibliography{bibliography}
\end{document}